\documentclass[conference]{newIEEEtran}
\pdfoutput=1

\usepackage{graphicx}
\usepackage{subfigure} 
\usepackage{amsmath}   

\usepackage{units}
\usepackage{accents}

\def\d{\textrm{d}}
\def\eq#1{\begin{equation}#1\end{equation}}
\def\vc#1{\accentset{\rightharpoonup}{#1}}
\def\uv#1{\hat{#1}}
\def\uG{\ensuremath{\mu\text{G}}}
\def\etal{\textit{et al.}}

\begin{document}

\title{UHECR propagation in the Galactic Magnetic Field}

\author{\authorblockN{Serguei Vorobiov\authorrefmark{1}\authorrefmark{2},
Mustafa Hussain\authorrefmark{1}, and
Darko Veberi\v{c}\authorrefmark{1}\authorrefmark{3}}
\\
\authorblockA{\authorrefmark{1}Laboratory for astroparticle physics,
University of Nova Gorica, Slovenia.}
\authorblockA{\authorrefmark{2}Email: sergey.vorobyev@p-ng.si}
\authorblockA{\authorrefmark{3}J.\ Stefan Institute, Ljubljana,
Slovenia}}

\maketitle

\begin{abstract}
Extensive simulations of the ultra-high energy cosmic ray (UHECR) propagation 
in the Galactic magnetic field (GMF) have been performed, and the results are presented.
The use of different available models of the large-scale GMF and/or primary 
particle assumptions leads to distinctly different deflection patterns 
of the highest energy cosmic rays (CR). The lensing effects of the Galactic field 
modify the exposure of an UHECR experiment to the extragalactic sky.
To quantify these effects for the Pierre Auger experiment, 
we performed a correlation analysis of the simulated cosmic ray event
samples, backtracked from
the Earth to the Galactic border, with the active galactic nuclei (AGN) 
from the 12th edition of the V\'eron-Cetty and V\'eron catalogue.
Further forward-tracking studies under plausible UHECR sources scenarios 
are needed to allow for direct comparison with the observed correlation 
between the nearby AGN and the highest energy Auger events.
\end{abstract}

\section{Introduction}

Magnetic fields in the Milky Way and other galaxies are investigated
by means of measurements of the Faraday rotation of the polarized light 
from pulsars and extragalactic sources, and through detection 
of synchrotron radiation emitted by relativistic electrons~\cite{beck,widrow}.  
The observations of spiral galaxies like ours suggest 
that there is a large-scale field, that in the first approximation 
follows the spiral arm structure, and is parallel to the galactic
disk. In the vicinity of the Sun such large-scale field 
is pointing approximately in direction of galactic latitude
$\ell=80^\circ$. The spiral field extends above and below 
the galactic disks on a kiloparsec scale and forms a kind of halo, 
which is confirmed by observations of the galaxies that are seen edge-on.
A random Galactic magnetic field, of the strength similar 
to that of the regular component, has also been observed~\cite{beck}. 
The coherence length of the turbulent field is at most of the order 
of several tens of parsecs. The average strength 
of the total magnetic field is about $6\,\mu{\rm G}$ 
near the Sun, and increases towards the Galactic Center region. 
The observations of the Galactic diffuse soft X-rays 
indicate on a possible CR and gas pressure driven wind, which 
would transport magnetic fields from the disk to the 
halo~\cite{everett,breitschwerdt}.

The observed Faraday rotation measures (RM) allow to probe
the Galactic magnetic field in the Solar
proximity~\cite{han,brown,men}.
However, the limited data sample makes the interpretation of these
observations difficult. In addition, the RM measurements provide 
the line-of-sight convolution of the magnetic field with 
the thermal electron density, which in turn is not well known. 
New radio polarization observations, 
with better sensitivity and angular resolution are needed to obtain 
more conclusive picture of the magnetic field distribution
in the Milky Way. A significant improvement of the data will be
provided by new-generation radio telescopes. 
One of such instruments, the Low Frequency Array (LOFAR), 
is already under construction~\cite{lofar}. LOFAR will be in particular
able to trace the magnetic fields in halo regions of 
the Milky Way and other galaxies, through detection of a few meters 
radio synchrotron emission from low energy cosmic
rays~\cite{beckfuture}.\\ 

The Pierre Auger Observatory~\cite{auger} provides a new and independent way 
of studying cosmic magnetic fields, by collecting cosmic ray events at the extreme 
energies (above $\unit[10]{EeV} \equiv \unit[10^{19}]{eV}$) with
unprecedented statistics and data quality. Recently, the Pierre Auger Collaboration 
observed~\cite{augeragncorrelationSci,augeragncorrelationAPh} 
a significant anisotropy in the arrival directions of cosmic rays above 
$\simeq \unit[60]{EeV}$. These cosmic rays correlate over angular
scales less than $6^\circ$ with the locations of nearby 
($D < \unit[100]{Mpc}$) AGN from the 12th edition 
of the V\'eron-Cetty and V\'eron (VCV) catalogue~\cite{vcv}.
The Pierre Auger experiment has also detected~\cite{augerspectrumPRL} 
a strong steepening of the cosmic ray flux above $4 \times 10^{19}$~eV.
Both observations are consistent with the standard scenarios 
of the UHECR production in the extra-galactic astrophysical 
acceleration sites and thus represent the important step 
towards the ``charged particle astronomy''. 

The final way the CR astronomy will be done
will strongly depend on the primary mass composition 
of the highest energy cosmic rays, since the magnetic deflections 
scale in proportion to the CR atomic number $Z$. The elaboration 
of relevant analysis methods requires detailed investigation 
of the UHECR propagation in cosmic magnetic
fields, and first of all in the Galactic field. The different aspects 
of the UHECR propagation in the GMF have been extensively 
covered in the literature, the non-exhaustive list of previous
results can be found 
in~\cite{stanev,harari,alvarez,tinyakov,prouza,kachelriess,takami}.
We present in this paper results of our own studies,
performed in the light of the observed AGN correlation.
We used the standard method of CR backtracking (see
Sec.~\ref{s:Backtracking}). Three distinctive large-scale field models
have been chosen (Sec.~\ref{s:GMFModels}). A large number 
of CR events has been simulated using energy spectrum, arrival
direction distribution, and mass composition described in 
Sec.~\ref{s:UHECRParameters}, and propagated under the assumed GMF models. 
Resulting magnetic deflections are presented 
in Sec.~\ref{s:MagneticDeflections}. Modification 
of the extragalactic exposure due to the large-scale GMF 
lensing effects is discussed in Sec.~\ref{s:EGExposureEffects}, 
followed by conclusions.

\begin{figure}[!h]
\centering
\includegraphics[width=0.4\textwidth]{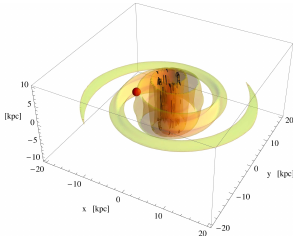}
\caption{The spiral disk field in the HMR models~\cite{harari}. 
The volume of galactic region with magnitude 
of magnetic field larger than $1\,\uG$ is shown in
yellow-green. The position of the Solar system is denoted by a red
sphere.}
\label{fig:HMRModel3DView}
\end{figure}

\section{CR Backtracking in the Galactic field}
\label{s:Backtracking}

Since the energy loss length in the local Universe largely exceeds the
size of the Galaxy (see, for example, \cite{achterberg,bhattacharjee,cronin}), 
we can neglect the energy losses when considering the propagation of the
ultra-high energy cosmic rays through the Milky Way. This
makes it possible to inverse the problem and, instead of following the
trajectory of a particle from the Galactic halo border till the
arrival on Earth from a certain direction, to propagate the
corresponding anti-particle from Earth in the same direction
(backtracking method). The trajectory of a UHE cosmic ray is then
obtained by integration of the equations of motion of an
ultra-relativistic particle with the energy $E$ and charge $q$ 
in a quasi-static magnetic field $\vc{B}$:
\eq{
\frac{\d\vc{r}}{\d ct} = \uv{v},
\qquad
\frac{\d\uv{v}}{\d ct} = \frac{qc}{E}\,\uv{v}\,\times\vc{B} \, ,
\label{EquationsOfMotionForm2}
}
where $\uv{v} = \vc{p}/|\vc{p}| \, , \, \vc{p} = E\d\vc{r} / c^2\d t$,
is the unit vector parallel to the velocity of the cosmic ray.
Since in the highly structured Galactic magnetic field the field
strength can vary significantly along the cosmic ray trajectory, we
have implemented for the integration of the equations of motion
\eqref{EquationsOfMotionForm2} the Runge-Kutta $5^\text{th}$
order scheme (RK5) with the adaptive step size control 
~\cite{numerical}. To avoid the ``numerical dissipation'' of 
the cosmic ray energy $E$ (the effect studied in detail 
in~\cite{armengaud}), we imposed the energy conservation during 
the propagation steps, by preserving the absolute value of the 
particle's velocity vector. 

The integration accuracy, represented by the accepted truncation 
error in the RK5 technique, has been optimized correlating 
the backtracked directions with the selected astrophysical objects.
An accuracy level of $10^{-6}$ has been adopted, for which
the changes in the backtracked directions become negligible with 
respect to the chosen step in angular separation $d_{\text{max}}$ 
from the catalogue objects (see the section~\ref{s:ScanParameters}).

The cosmic ray trajectory has been followed till the galactocentric 
distance of $\unit[20]{kpc}$ (the Galaxy ``border''), beyond which 
the field strength is supposed to be negligibly small.

\section{Considered large-scale GMF models}
\label{s:GMFModels}

A strong large-scale magnetic field with a stationary or oscillating 
configuration can be generated from a weak seed field as a result
of the non-uniform (differential) rotation of the galactic gas with 
strong turbulent motions~\cite{widrow}. The regular field structure 
obtained by this so-called dynamo mechanism can be described by modes 
of different azimuthal symmetry in the disk, and vertical 
symmetry perpendicular to the disk plane. The strongest mode is
the spiral disk field with either $\pi$ (\textit{bisymmetric}, or BSS) 
or $2\pi$ (\textit{axisymmetric}, or ASS) symmetry. Other weaker field modes 
can also be excited.  As for the vertical symmetry, 
the large-scale galactic fields can either keep the direction while traversing 
the disk plane (be of~\textit{even parity}), either change it 
to the opposite (\textit{odd parity} modes). Both symmetry properties 
are reflected in the toponym of the large-scale field models. Combined 
with the spiral structure type, 4 possible spiral field patterns are denoted 
\textbf{BSS-S, BSS-A, ASS-S, ASS-A}, where -S (symmetric) stands 
for the even parity, and -A (antisymmetric) -- for the odd parity.

We have chosen amongst many available large-scale GMF models three
typical ones with distinct qualitative differences. 
Two models are the spiral disk field models of~\textit{bisymmetric even
parity}, and \textit{axisymmetric odd parity} from the paper~\cite{harari}
by Harari, Mollerach and Roulet (HMR). These models 
(see Fig.~\ref{fig:HMRModel3DView}) represent a version of the Stanev 
model~\cite{stanev}, smoothed out in order to avoid the field discontinuities. 

The third model is a modification of the model \cite{prouza}
by Prouza and \v{S}m\'ida (PS), that has been proposed by 
Kachelrie\ss~\etal~\cite{kachelriess}. In addition to the 
\textit{bisymmetric even parity} spiral field, of the structure
similar to the one in the HMR BSS\_S model, it features 
two additional large-scale halo GMF components: 
\textit{toroidal} azimuthal field above and below the Galactic disk, 
and \textit{poloidal} (dipole) field.

\section{Parameters adopted for simulated CR events}
\label{s:UHECRParameters}

We simulated a large number (stated in 
Tab.~\ref{t:DeflectionDistrPercentiles} 
and~\ref{t:log10PminDistrPercentiles}) 
of cosmic ray events with the parameters below.  

\subsection{Energy spectrum}

The energy values of the simulated events were bounded 
between \unit[40]{EeV} and \unit[150]{EeV}, and distributed
according to the Auger measurement~\cite{augerspectrumPRL}, 
as a power law $E^{-4.2}$. 

\subsection{Angular distribution}

Two options for the angular distribution of the events have been 
considered: 
1) Uniform distribution over the whole sky, and 
2) Distribution according to the exposure of 
the Auger Surface Detector (SD) for zenith angles 
$\theta_\text{z} \le 60^\circ$. 
For the considered energies the Auger SD acceptance area is
saturated, 
which produces a simple analytic dependence of the exposure 
upon declination~\cite{sommers}.

\subsection{Primary mass composition}

4 pure compositions have been considered:
protons ($Z=1$), carbon ($Z=6$), silicon ($Z=14$), 
and iron ($Z=26$) nuclei. 

\begin{figure*}[!ht]
\centerline{\subfigure[Bisymmetric even parity spiral field]
{\includegraphics[width=0.5\textwidth]{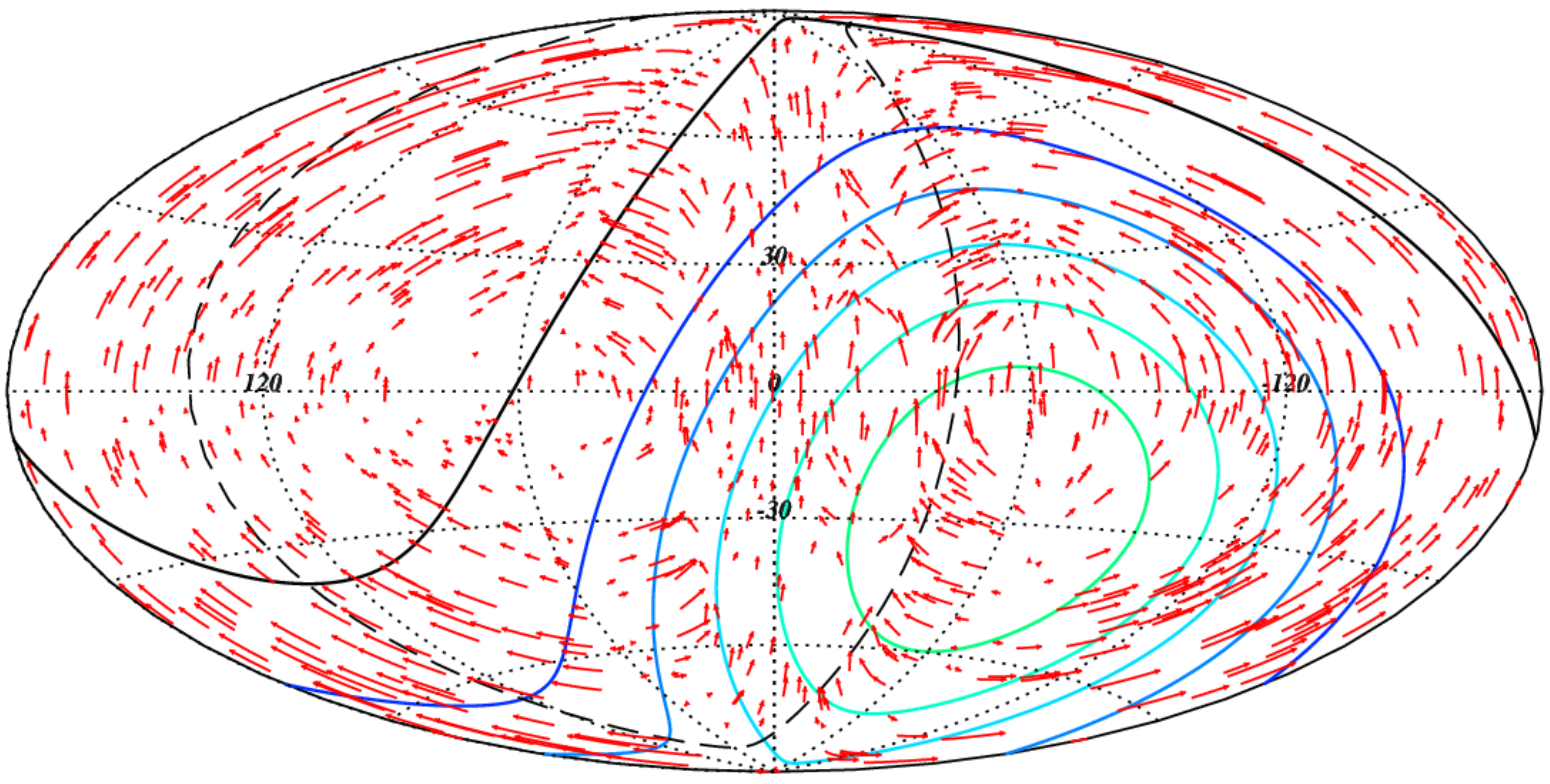}
\label{fig:hmrbsss_defldir}}
\hfil
\subfigure[Axisymmetric odd parity spiral field]
{\includegraphics[width=0.5\textwidth]{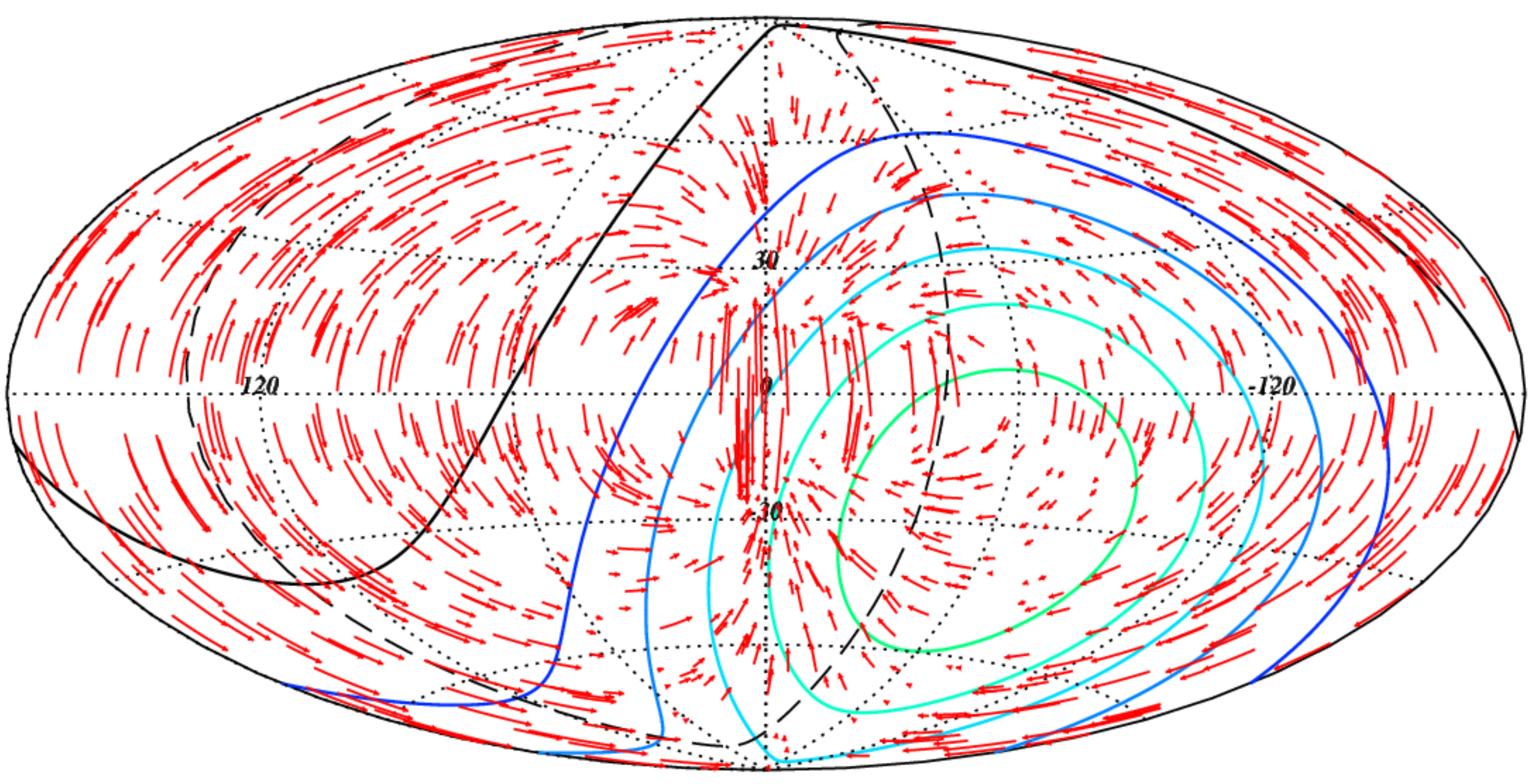}
\label{fig:hmrassa_defldir}}}
\caption{Cosmic ray deflection patterns expected for \emph{protons} 
under HMR models~\cite{harari}.
The deflections shown represent the arcs of
the great circle joining the simulated arrival direction on Earth, and
the backtracked one at the Galaxy border (indicated by an
arrow head). To
visualize deflections, every $100^{th}$ event has been drawn from
the large simulated event data sample described in the 
section~\ref{s:UHECRParameters}.  The dashed line denotes the
supergalactic plane. The thin solid lines show the equal integrated
Auger SD exposure sky regions within the detector field of view 
for zenith angles $\theta_\text{z} \le 60^\circ$, delimited by the
black solid line.}
\label{fig:hmrmodels_defldir}
\end{figure*}

\section{Magnetic deflections of cosmic rays}
\label{s:MagneticDeflections}

\begin{figure}[!h]
\centering
\includegraphics[width=0.5\textwidth]{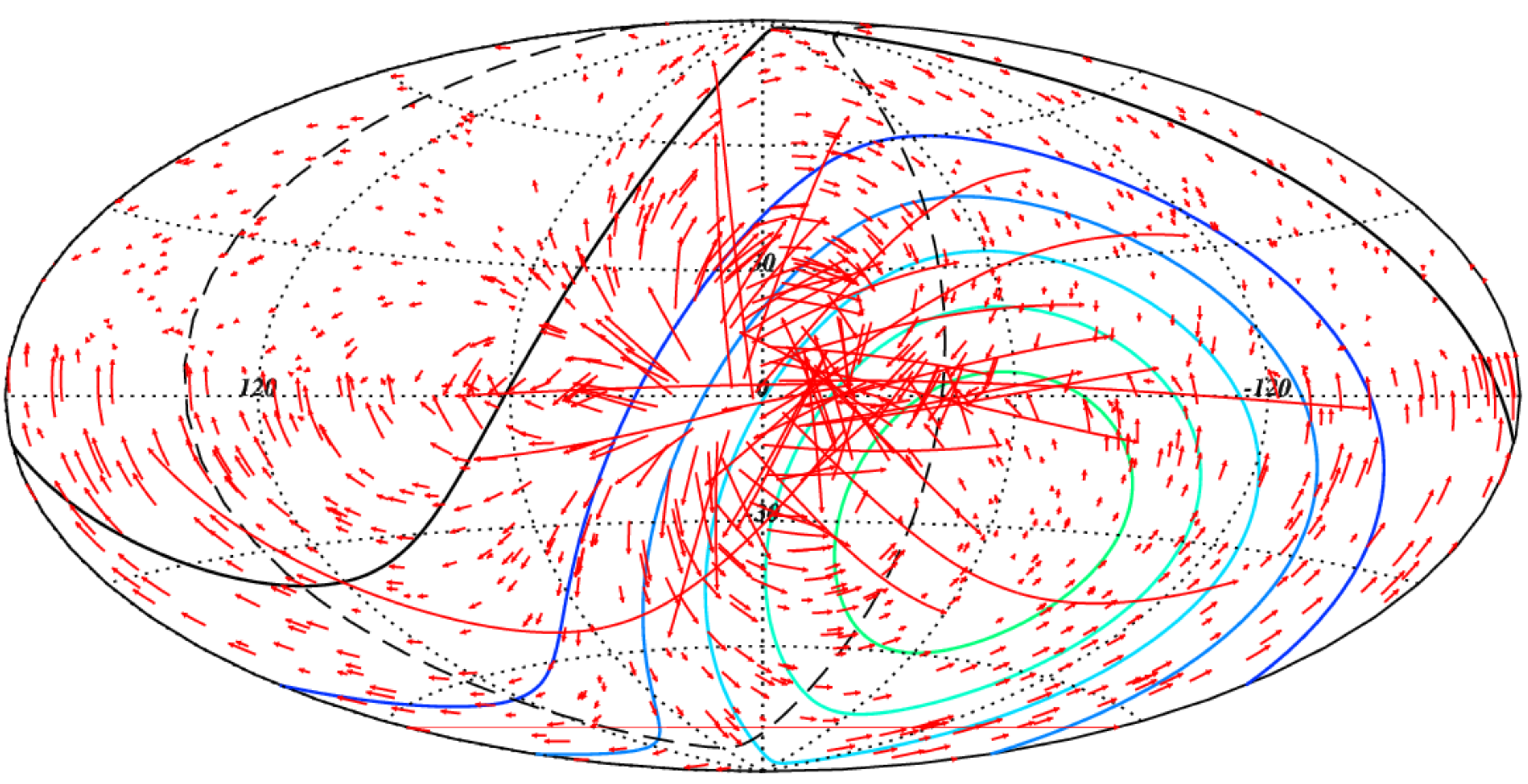}
\caption{\emph{Idem} as Fig.~\ref{fig:hmrmodels_defldir} for the PS model~\cite{prouza} 
version by Kachelrie\ss~\etal~\cite{kachelriess}.}
\label{fig:pskstmodel_defldir}
\end{figure}

The cosmic ray deflection patterns expected from the three 
selected GMF models are very different, as it can be seen from 
Figs.~\ref{fig:hmrmodels_defldir} and~\ref{fig:pskstmodel_defldir}
on the example of primary \emph{protons}.  
Except for the directions towards the Galactic Center, where the
character of deflections is complex due to the multiple field
reversals, there are trends, typical for each
model. In the bisymmetric even parity spiral disk field, there is a
flow in the direction of the Galactic North Pole, i.e. cosmic rays
in general arrive at the Galactic halo border from higher Galactic
latitudes than observed on Earth.  In the odd parity field, the
backtracked arrival directions in each Galactic hemisphere are
shifted to the poles. This field configuration would obscure the Earth
sky from the cosmic rays coming from the directions close to the
Galactic plane. The presence of dipole and toroidal fields
makes the deflection pattern even more complex (see Fig.~\ref{fig:pskstmodel_defldir}).

\begin{figure}[!h]
\centering
\includegraphics[width=0.42\textwidth]
{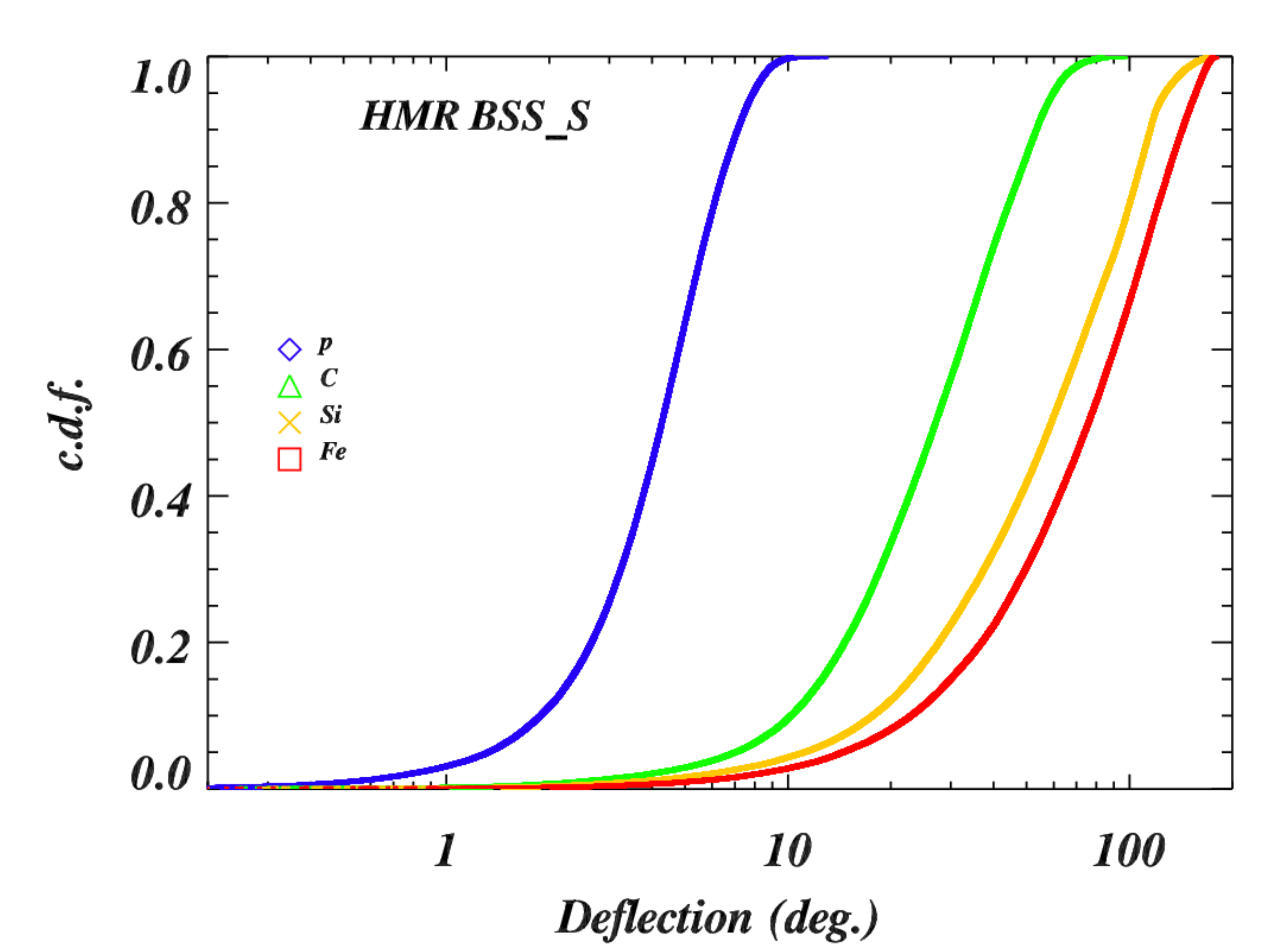}
\caption{C.d.f. of the deflection angle
values (in $^\circ$), expected in the HMR BSS\_S model \cite{harari}
for the four assumed mass compositions.}
\label{f:DeflectionDistrCDF}
\end{figure}

\begin{table}[!h]
\centering
\begin{tabular}{l|l||c|c|c|c}
\multicolumn{2}{c||}{\textbf{Primary}}&\multicolumn{4}{c}{\textbf{Uniform $4\pi$ exposure:}}
\\
\hline
 & Z & $N_{events}$ & $\vartheta_{50\%}$ & $\vartheta_{68.27\%}$ & $\vartheta_{95.45\%}$
\\
\hline
\multicolumn{6}{c}{}
\\
\multicolumn{6}{c}{\textbf{HMR bisymmetric even parity model}}
\\
\hline
 p &   1  & 100000 &	 4.3 &     5.3 &     8.0\\
 C &   6  & 100000 &	27.1 &    36.5 &    60.7\\
Si &  14  & 100000 &	59.1 &    82.6 &   129.1\\
Fe &  26  & 100000 &	75.9 &   102.7 &   157.7\\
\multicolumn{6}{c}{}
\\
\multicolumn{6}{c}{\textbf{HMR axisymmetric odd parity model}}
\\
\hline
 p &   1  & 100000 &     5.4 &     6.6 &    10.9\\
 C &   6  & 100000 &    31.5 &    40.2 &    77.9\\
Si &  14  & 100000 &    75.9 &    97.6 &   139.2\\
Fe &  26  & 100000 &    84.0 &   100.4 &   138.0\\
\multicolumn{6}{c}{}
\\
\multicolumn{6}{c}{\textbf{PS model version by Kachelrie\ss~\etal}}
\\
\hline
 p &   1 &   20000 &	  3.0 &     4.3 &    20.8\\
 C &   6 &   20000 &	 16.7 &    22.9 &    64.7\\
Si &  14 &   20000 &	 37.6 &    47.2 &    85.7\\
Fe &  26 &   20000 &    57.5 &    70.9 &   118.5\\
\end{tabular}
\caption{Magnetic deflections(in $^\circ$) at the indicated
percentiles of the c.d.f. (shown on Fig~\ref{f:DeflectionDistrCDF}
for one of the three GMF models).}
\label{t:DeflectionDistrPercentiles}
\end{table}

The cumulative distributions (c.d.f.) of resulting deflections 
for the HMR BSS\_S model are shown on
Fig.~\ref{f:DeflectionDistrCDF}. 
The deflection values for the three considered GMF models 
are summarized in the table~\ref{t:DeflectionDistrPercentiles}, by
means of percentiles at 50\% (median), 68.27\%, and 95.45\% of
the c.d.f. Neglecting the distribution tails,
one can see that the deflection values scale rather well with the
atomic number $Z$ of primary nuclei. Despite the additional halo
fields in the PS model version, the higher overall normalization of
the disk field strength in the HMR models, and its more important halo
extension~\cite{harari,prouza,kachelriess} are the
reasons why in average the deflections for the HMR models are larger.

It is instructive to compare the obtained UHECR deflections 
(see Fig.~\ref{f:DeflectionDistrCDF}) with those expected for turbulent 
field with a coherence length $l_{\text{coh}}$. In such a
case, on a distance $d \gg l_{\text{coh}}$ a cosmic ray 
will propagate in the random walk regime, with the resulting 
r.m.s. deflection angle given by~\cite{achterberg,bhattacharjee}

\eq{
\vartheta_{random}\,[^\circ] \sim 1.13^\circ\,Z\,
  \frac{\sqrt{d\,[\unit{kpc}]}\,\sqrt{l_{\text{coh}}\,[\unit[100]{pc}]}\,B\,[\mu{\rm G}]}
       {E\,[\unit[10]{EeV}]},
\label{DeflRandomField}
}

where the effect of dynamical friction~\cite{achterberg} is taken into 
account. The path $d$ can be roughly estimated using the simulated events 
(see Sec.~\ref{s:UHECRParameters}). For the HMR BSS\_S field configuration, 
the median distance $d_{50\%}$ traveled till the Galaxy border 
is \unit[21.0]{kpc} for protons, and \unit[28.3]{kpc} for iron nuclei,
while the respective values for the two primaries corresponding to 
95.45\% of the c.d.f. are \unit[28.1]{kpc} and \unit[70.3]{kpc}.
Under reasonable assumption of the equal field strength for
the large-scale and turbulent components, an r.m.s. turbulent field
strength is expected to decrease with a distance to the Galactic
plane at a kiloparsec scale, which is much smaller than the
above mentioned $d$ values. Hence, for the energies of interest 
and for most of the directions in the sky the UHECR deflections 
from the random Galactic field component \eqref{DeflRandomField} 
will be considerably smaller than the ones from the large-scale 
GMF (see also~\cite{tinyakovrandom}). Therefore, the former can be 
neglected at the first approach, though a realistic simulation 
of the UHECR propagation in the Galactic magnetic field 
has to take into account the turbulent field.

\section{GMF effects on the extragalactic exposure}
\label{s:EGExposureEffects}

\begin{figure}[!h]
\centering
\includegraphics[width=0.5\textwidth]
{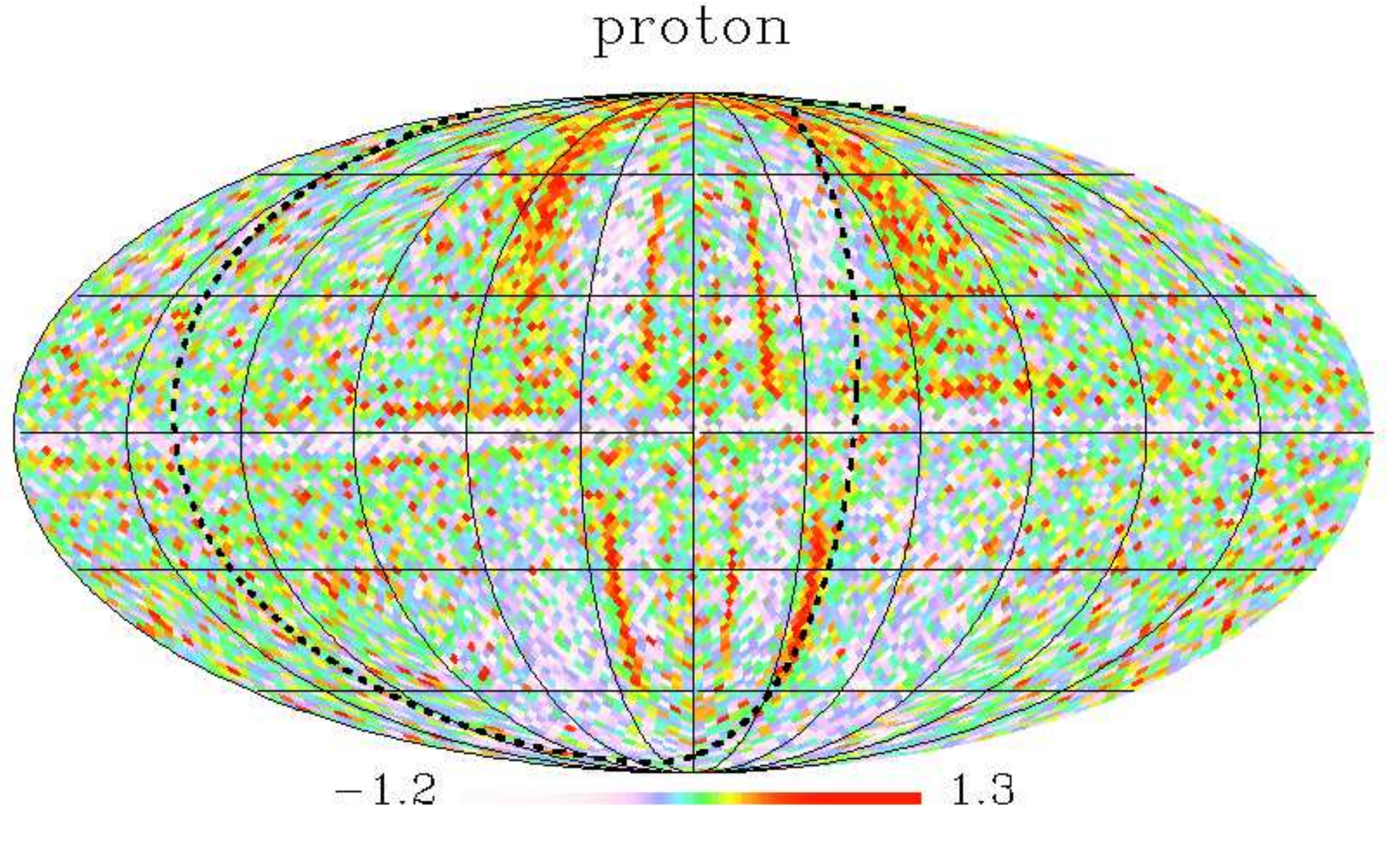}
\includegraphics[width=0.5\textwidth]
{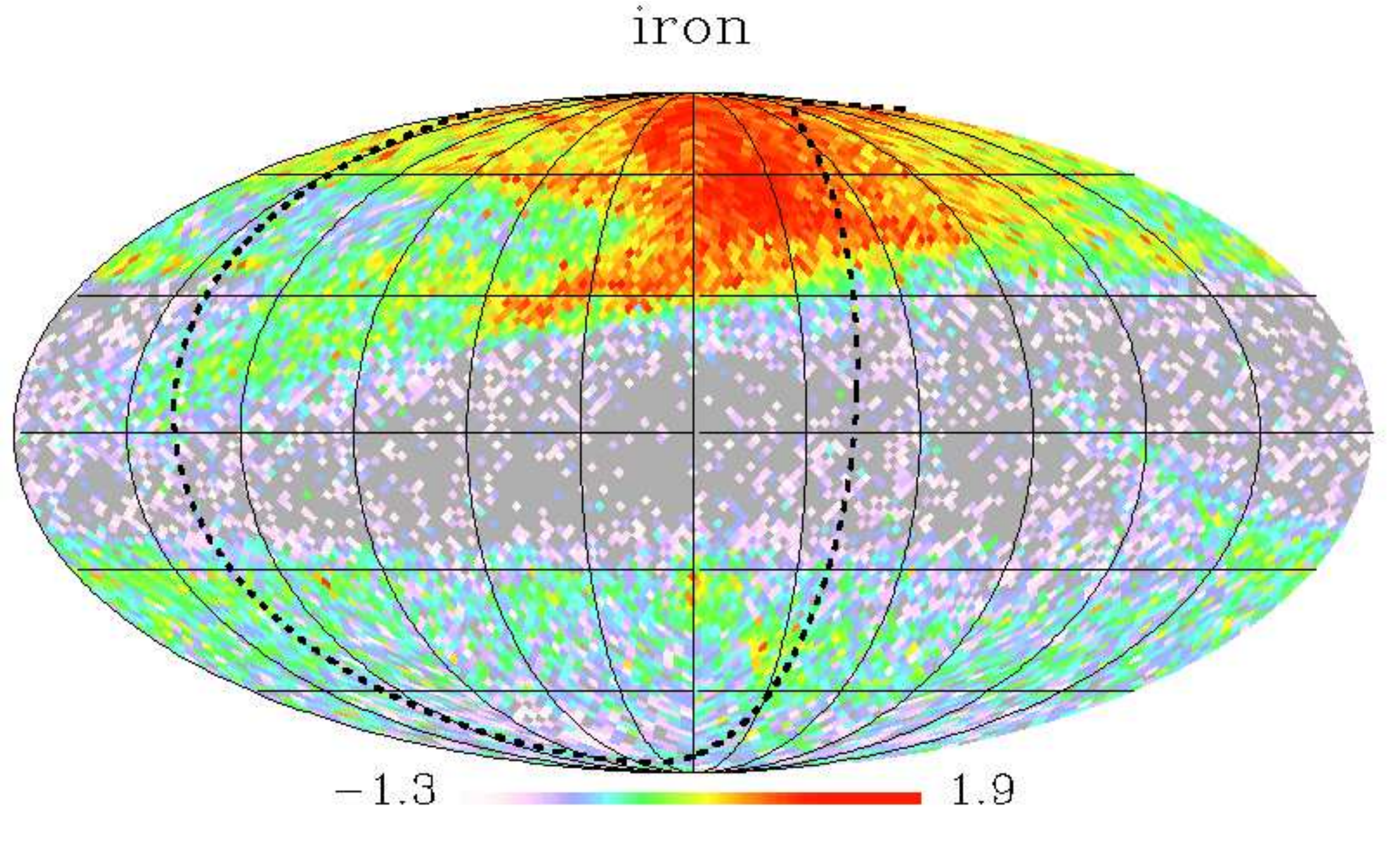}
\caption{The sky distribution of the $\log_{10}$ of the
ratio of the number of events at the Galaxy border
to the one on Earth, for the HMR
\textit{bisymmetric even parity} model~\cite{harari}. The gray color represents 
the regions on the sky, to which the Earth detectors are 
(almost) blind, like the one towards the Galactic Center
on the lower plot. 
The arrival direction distribution on Earth is isotropic. 
The assumed cosmic ray primary type is indicated above each map. 
The thick dashed line denotes the
supergalactic plane.}
\label{f:EGExposureMapHMRBSSS}
\end{figure}

\begin{figure}[!h]
\centering
\includegraphics[width=0.5\textwidth]
{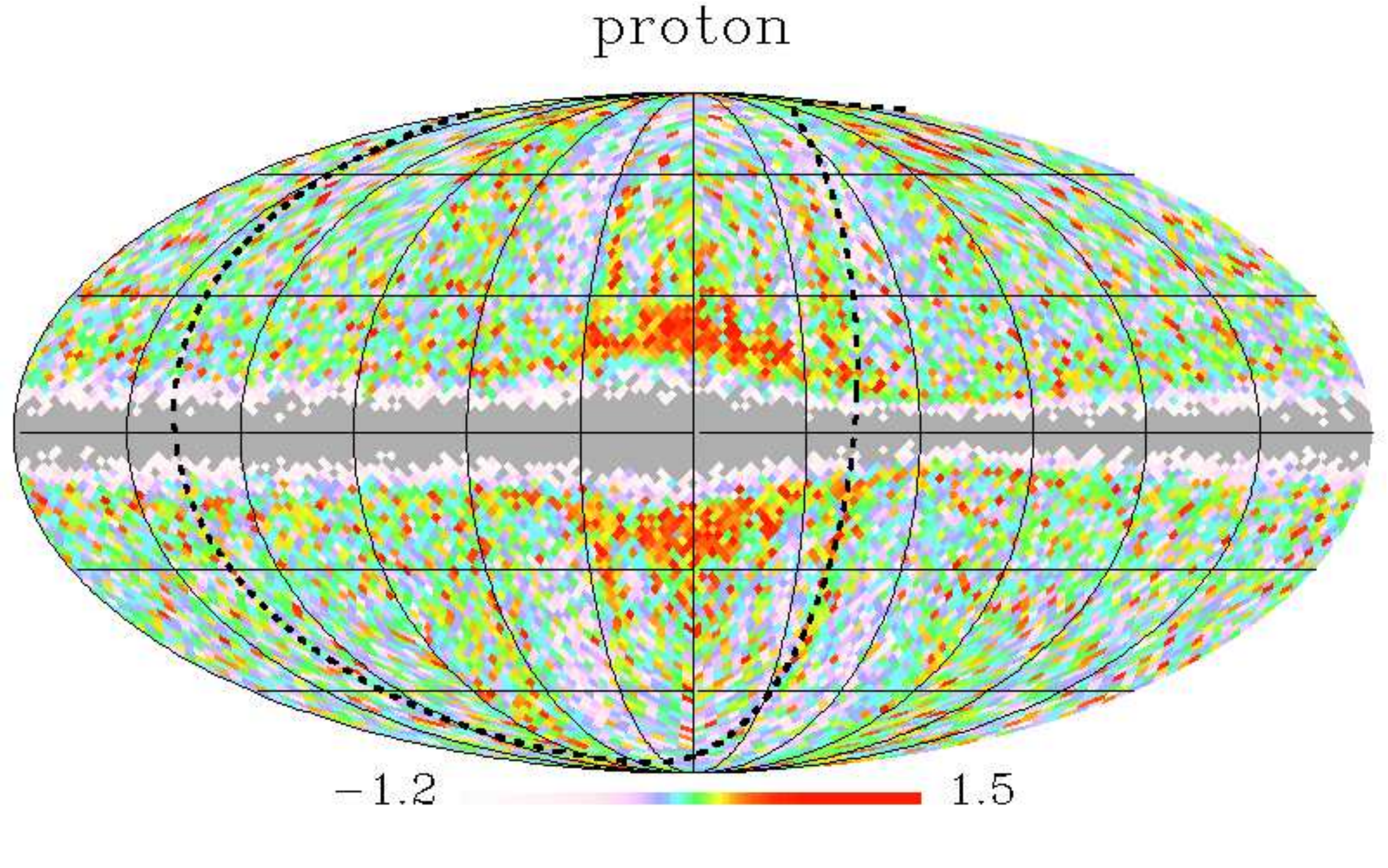}
\includegraphics[width=0.5\textwidth]
{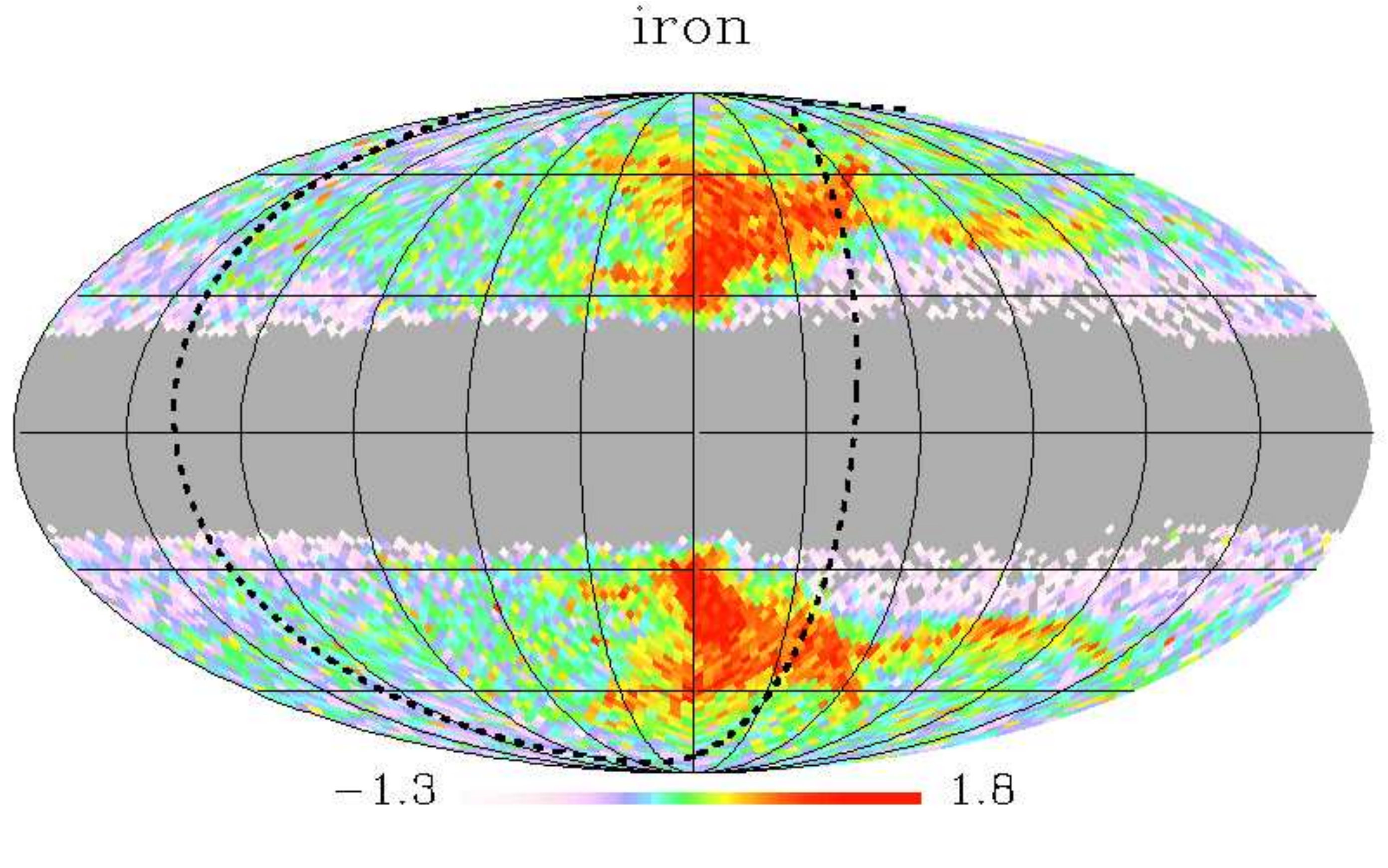}
\caption{\emph{Idem} as Fig.~\ref{f:EGExposureMapHMRBSSS} 
for the HMR \textit{axisymmetric odd parity} model~\cite{harari}.}
\label{f:EGExposureMapHMRASSA}
\end{figure}

\begin{figure}[!h]
\centering
\includegraphics[width=0.5\textwidth]
{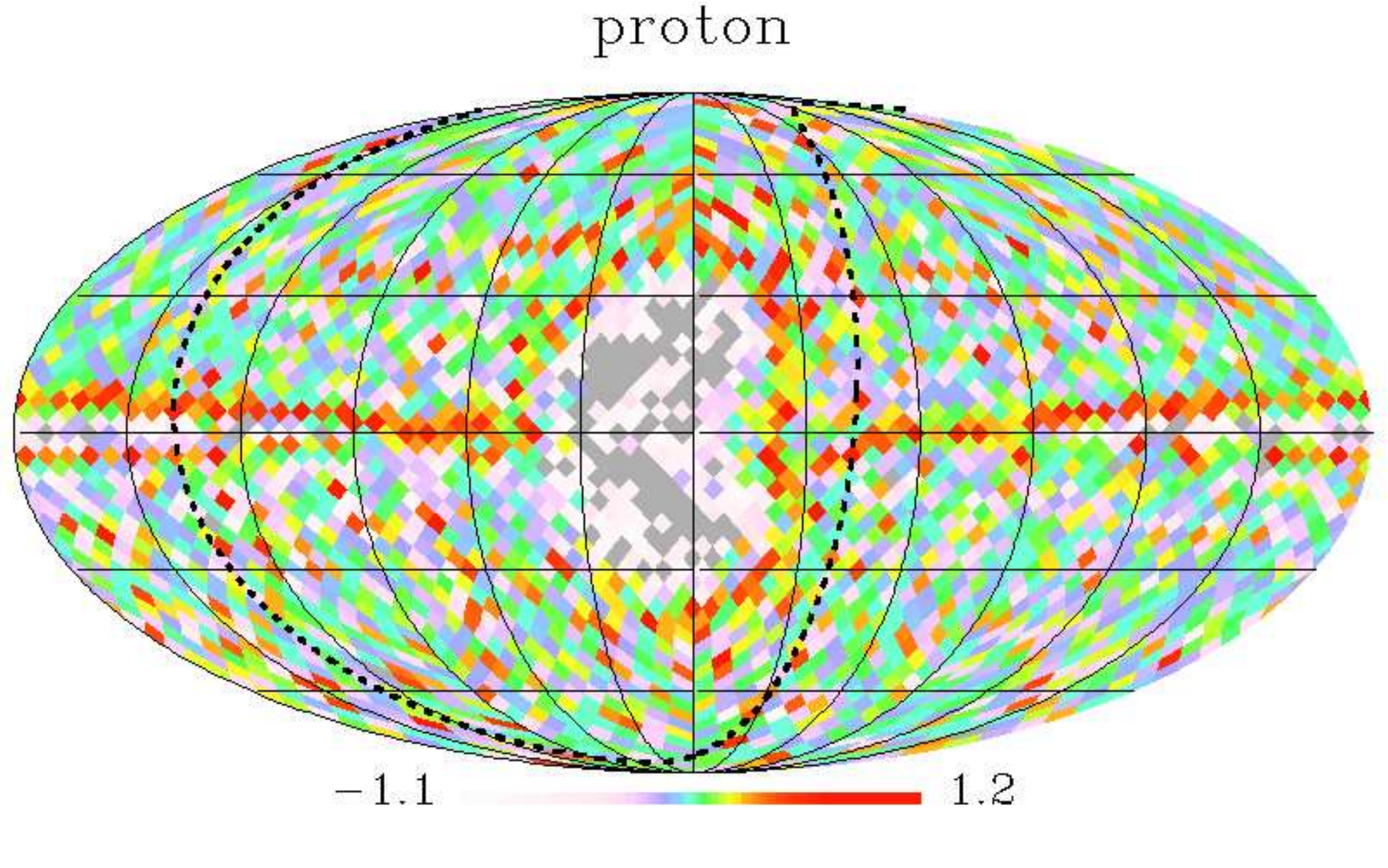}
\includegraphics[width=0.5\textwidth]
{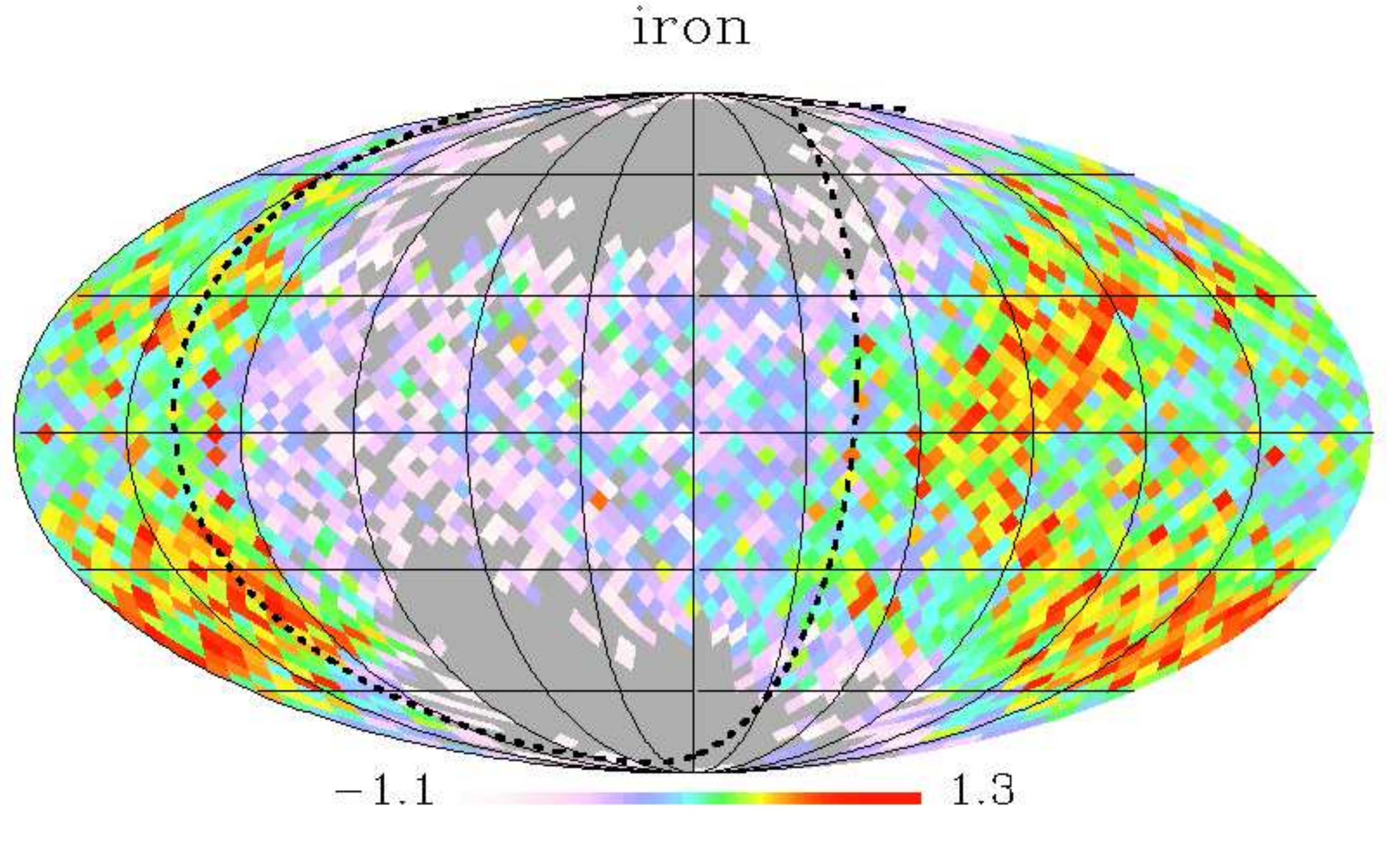}
\caption{\emph{Idem} as Fig.~\ref{f:EGExposureMapHMRBSSS} for the PS model~\cite{prouza} 
version by Kachelrie\ss~\etal~\cite{kachelriess}.}
\label{f:EGExposureMapPSKSTModel}
\end{figure}

In addition to magnetic deflections, (de-)focusing effects of the Galactic
magnetic field will modify the exposure of an experiment to the extragalactic
sky~\cite{harari,alvarez,kachelriess}. Though the Liouville theorem holds
and the isotropic cosmic ray flux outside the Galactic halo will remain 
isotropic on Earth, the correspondence between the arrival directions on
Earth, and the sources contributing to the flux on the halo border will be 
quite complex. To study how this correspondence is realized, we have used
the backtracked simulated events from the section~\ref{s:UHECRParameters}.
We mapped the decimal logarithm of the ratio of the number of events
on the halo border ($r = \unit[20]{kpc}$) to the one on Earth, using 
the HEALPix equal area celestial sphere pixelization~\cite{healpix}.
These maps, shown on Figures~\ref{f:EGExposureMapHMRBSSS},
\ref{f:EGExposureMapHMRASSA}, and~\ref{f:EGExposureMapPSKSTModel} 
allow to estimate the (de-)magnification effects of the 
large-scale field on the experimental exposure. 

Due to these effects, some regions of the
sky have increased probability to contribute to the cosmic rays
observed on Earth, and some other are disfavored.
The non-uniform character of the mapping strengthens in the case
of the heavy primary composition. There exist directions,
for which the Earth detectors are blind, i.e. even if there is
a strong UHECR source in such a ``discriminated'' sky
region, heavy nuclei from this source will not reach the Earth. 
The highly non-uniform mapping in the case of pure iron 
primary composition is demonstrated by the lower plots 
on Figs.~\ref{f:EGExposureMapHMRBSSS},~\ref{f:EGExposureMapHMRASSA},
and \ref{f:EGExposureMapPSKSTModel}.
If the primary cosmic rays above 40 EeV are iron nuclei, 
and the GMF structure is adequately described by one of the assumed
large-scale GMF models, the regions that would effectively contribute
to the cosmic ray flux on Earth represent only a small fraction
of the $4\pi$ solid angle.  

\subsection{Correlation scan using backtracked directions}
\label{s:ScanParameters}

To quantify these large-scale GMF effects on the mapping between 
the arrival directions on Earth and those at the Galaxy border
in the case of the Pierre Auger Observatory, we performed 
a correlation analysis using \textit{backtracked} arrival directions
of simulated events described in Sec.~\ref{s:UHECRParameters},
and the 694 AGN at redshift $z_{\text{max}} \le 0.024$ 
from the VCV catalogue. The simulated events have been divided sequentially
into samples with the same number of events (81) above
$\unit[40]{EeV}$ as in the Auger data.
The employed set and ranges of parameters were also identical to the ones 
from~\cite{augeragncorrelationSci,augeragncorrelationAPh}.
The lower threshold energy giving maximal correlation was scanned within 
a range $E_{{\text{min}}}\ge\unit[40]{EeV}$. The maximum redshift
$z_{\text{max}}$ scan was performed within a range $0 \le z_{\text{max}} \le 0.024$,
in steps of 0.001. The scan in maximum angular distance $d_{\text{max}}$ was made
within a range  $1^\circ \le d_{\text{max}} \le 8^\circ$, in steps of $0.1^\circ$.

\begin{figure}[!h]
\centering
\includegraphics[width=0.45\textwidth]
{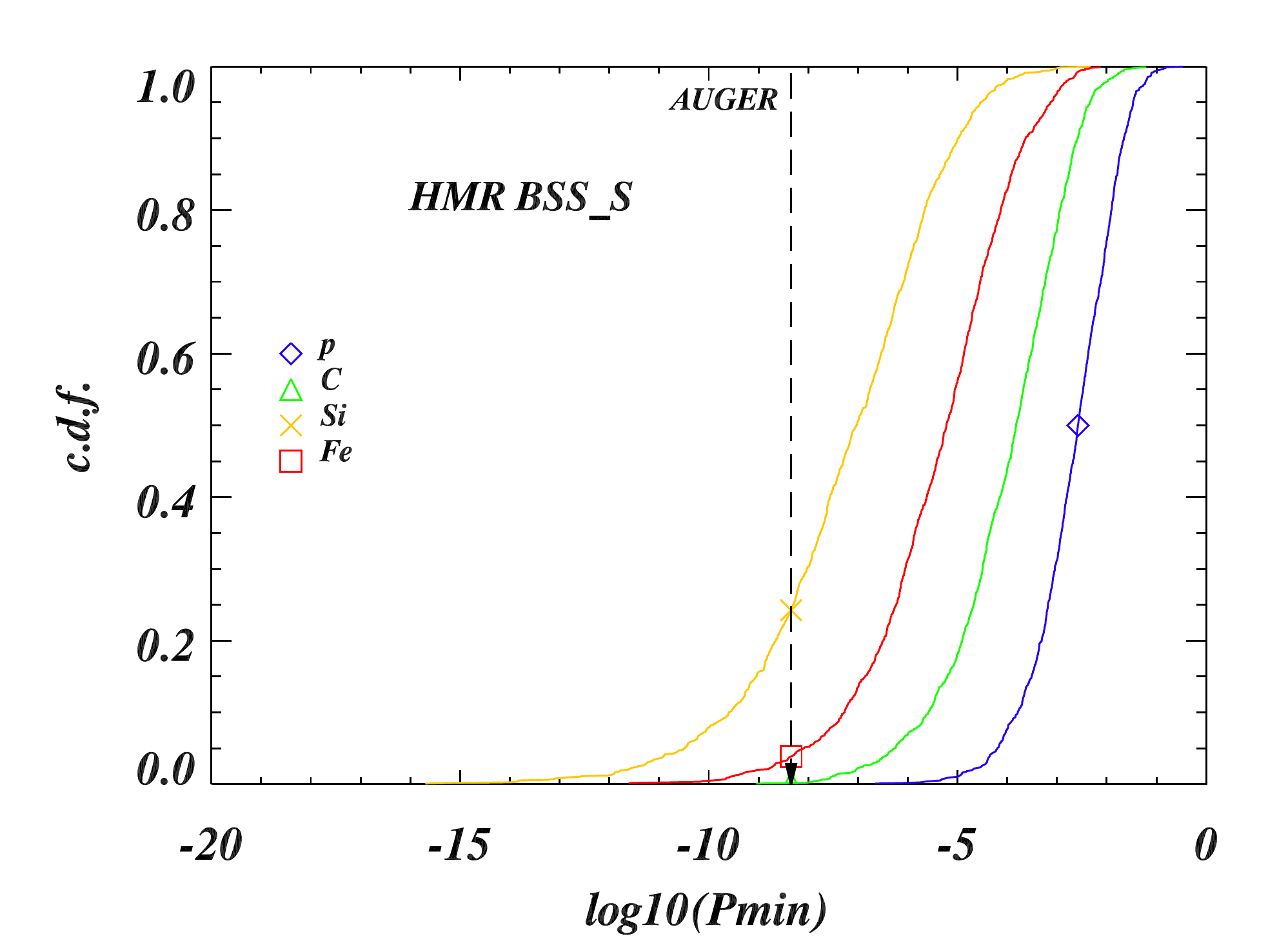}
\caption{Scan results for the
simulated event samples, backtracked under
the HMR \textit{bisymmetric even type} model~\cite{harari}.
The four cumulative distributions of $\log_{10}P_{\text{min}}$
correspond to primary assumptions of pure protons,
carbon, silicon, or iron nuclei.
The vertical dashed line indicates 
the value obtained from the scan of 
the Auger events~\cite{augeragncorrelationSci,
augeragncorrelationAPh} \textit{on Earth}.}
\label{f:PlotScanResMC_HMRBSSS}
\end{figure}

We will focus here on the minimum probability values, obtained during
the scan. The c.d.f. of decimal logarithm of the corresponding cumulative binomial
probability $P_{{\text{min}}}$ of reaching this level of correlation 
under isotropy for the HMR BSS\_S model are shown on 
Fig.~\ref{f:PlotScanResMC_HMRBSSS}. The level of the minimum
probability reached in the Auger data is also indicated. 
The results of the correlation scan for the three assumed large-scale
GMF models and four primary mass compositions are summarized in 
Table~\ref{t:log10PminDistrPercentiles}.

\begin{table}[!h]
\centering
\begin{tabular}{l|l||c|c|c|c}
\multicolumn{2}{c||}{\textbf{Primary}}&\multicolumn{4}{c}{\textbf{Auger SD exposure:}}
\\
\hline
 & Z & $N^{\textit{81}\ event}_{samples}$ & $f_{5\%}$ & $f_{50\%}$ & $f_{95\%}$
\\
\hline
\multicolumn{6}{c}{}
\\
\multicolumn{6}{c}{\textbf{HMR bisymmetric even parity model}}
\\
\hline
 p &   1  & 1148 &   -4.20 &   -2.58 &   -1.47\\
 C &   6  & 1148 &   -6.26 &   -3.82 &   -2.34\\
Si &  14  & 1147 &  -10.49 &   -7.03 &   -4.53\\
Fe &  26  & 1148 &   -8.10 &   -5.22 &   -3.13\\
\multicolumn{6}{c}{}
\\
\multicolumn{6}{c}{\textbf{HMR axisymmetric odd parity model}}
\\
\hline
 p &   1  & 1148 &   -4.48 &   -2.70 &   -1.57\\
 C &   6  & 1148 &   -8.69 &   -5.78 &   -3.70\\
Si &  14  & 1148 &  -12.56 &   -8.50 &   -5.30\\
Fe &  26  & 1148 &   -8.98 &   -5.89 &   -3.85\\
\multicolumn{6}{c}{}
\\
\multicolumn{6}{c}{\textbf{PS model version by Kachelrie\ss~\etal}}
\\
\hline
 p &   1 &     308 &   -4.46 &   -2.66 &   -1.45\\
 C &   6 &     308 &   -4.55 &   -2.55 &   -1.43\\
Si &  14 &     308 &   -5.21 &   -2.92 &   -1.72\\
Fe &  26 &     308 &   -4.46 &   -2.52 &   -1.31\\
\end{tabular}
\caption{Values of $\log_{10}P_{\text{min}}$
at the indicated percentiles of the c.d.f.,
for the assumed GMF models 
and primary compositions.}
\label{t:log10PminDistrPercentiles}
\end{table}

Since for the different GMF model/primary mass assumptions the backtracked directions 
correlate with the catalogue objects in particular privileged regions in the sky, 
our scan results depend strongly on those assumptions. 
Under some configurations, the backtracked directions lie near the supergalactic
plane, where the catalog object density is higher than in average,
which increases the correlation probability. For the PS model~\cite{prouza}
version by Kachelrie\ss~\etal~\cite{kachelriess}, the scanned probability
minimum is significantly less deep than for the two other 
spiral-field-only models for any assumed primary mass composition, 
except for protons, where one obtains nearly the same level of
correlation for all models.  
The model with additional halo field components is clearly less
compatible with the observed correlation with the nearby 
AGN~\cite{augeragncorrelationSci,augeragncorrelationAPh}, 
unless the UHECR flux contribution from these objects (or
objects with similar spatial distribution) is highly non-uniform
(for example, if there are only few powerful sources, 
one of which is located by chance in
one of those regions in the sky, where the large-scale field 
``enhances'' the Auger exposure).

The presented approach has therefore a potential of providing constraints 
that can be used to discriminate between the GMF models and/or the primary charge.
It is now being complemented by a forward-tracking
of cosmic rays to the Earth, for a number of plausible
UHECR sources scenarios.   

\section{Conclusions}
\label{s:Conclusions}

We have studied the propagation of highest energy cosmic rays
in the Galactic magnetic field, for different assumptions on the large-scale
field structure and/or primary cosmic ray mass composition.
The knowledge of the field distribution in the Milky Way is limited
by the sensitivity and angular resolution of the present-day instruments. 
The Pierre Auger Observatory data provide the measurements of the primary cosmic ray 
properties with unprecedented level of collected statistics and reconstruction 
accuracy. This allows a complementary and independent way of probing 
the Galactic field structure. The observed correlation
between the highest energy Auger events and the nearby AGN 
may give additional hints about the GMF amplitude and orientation.\\

We have investigated deflection patterns of cosmic rays above $\unit[40]{EeV}$,
underwent during their propagation in the Galactic field under three distinctively 
different GMF models. The exact deflection value in the regular component 
of the field depends strongly on the arrival direction on Earth of a cosmic ray,
with the corresponding position angle of deflection differing from one assumed
field distribution to another. In addition, the spectrometric character of deflections
of the UHECR in the large-scale Galactic magnetic field gives rise to the aligned
structures of events coming from a UHECR source (thread-like multiplets), that
can be used to discriminate between GMF models (see~\cite{multiplets}
and the references therein).\\

Magnetic deflections and (de-)focusing effects of the Galactic magnetic field
modify the exposure of an experiment to the extragalactic sky. Our studies
show that the UHECR picture observed on Earth is sensitive
to the Galactic field distribution and the primary cosmic ray composition. 
Though the reconstruction of the field is easier in the case of light
primary mass composition, in the case of heavy nuclei the lensing effects 
of the Galactic field on the exposure bring stronger constraints 
on the list of potential UHECR source candidates. The presented analysis 
of the correspondence between the arrival direction distributions on Earth
and at the Galactic border, and of matching of the latter with the VCV
catalogue objects will be complemented by the forward-tracking of cosmic rays 
for a number of plausible UHECR sources scenarios. This will allow for direct
comparison with the observed AGN correlation.

\section*{Acknowledgements}
We acknowledge the discussions and cross-checks of results
with E.~M.~Santos at the initial stage of the presented work.

\end{document}